\documentclass{bmcart}

\usepackage{amsthm,amsmath}
\usepackage{braket}
\usepackage[utf8]{inputenc} %unicode support
\usepackage{graphicx}
\usepackage{chngcntr} % 用于改变计数器
\startlocaldefs
\endlocaldefs

\begin{document}

%%% Start of article front matter
\begin{frontmatter}

\begin{fmbox}
\dochead{Research}

%%%%%%%%%%%%%%%%%%%%%%%%%%%%%%%%%%%%%%%%%%%%%%
%%                                          %%
%% Enter the title of your article here     %%
%%                                          %%
%%%%%%%%%%%%%%%%%%%%%%%%%%%%%%%%%%%%%%%%%%%%%%

\title{Enhanced high-dimensional teleportation in correlated amplitude damping noise by weak measurement and environment-assisted measurement}

%%%%%%%%%%%%%%%%%%%%%%%%%%%%%%%%%%%%%%%%%%%%%%
%%                                          %%
%% Enter the authors here                   %%
%%                                          %%
%% Specify information, if available,       %%
%% in the form:                             %%
%%   <key>={<id1>,<id2>}                    %%
%%   <key>=                                 %%
%% Comment or delete the keys which are     %%
%% not used. Repeat \author command as much %%
%% as required.                             %%
%%                                          %%
%%%%%%%%%%%%%%%%%%%%%%%%%%%%%%%%%%%%%%%%%%%%%%

\author[
  addressref={aff1},                % id's of addresses, e.g. {aff1,aff2}
  %corref={aff1},                      % id of corresponding address, if any
% noteref={n1},                        % id's of article notes, if any
  %email={carlos.sabin@uam.es}   % email address
]{\inits{X.X.}\fnm{Xing} \snm{Xiao}}
\author[
  addressref={aff1},                  % id's of addresses, e.g. {aff1,aff2}
  corref={aff1},                      % id of corresponding address, if any
% noteref={n1},                        % id's of article notes, if any
  email={lu.tianxiang@foxmail.com}   % email address
]{\inits{T.X.L.}\fnm{Tian-Xiang} \snm{Lu}}
\author[
  addressref={aff2},                   % id's of addresses, e.g. {aff1,aff2}
  corref={aff2},                       % id of corresponding address, if any
% noteref={n1},                        % id's of article notes, if any
  email={liyanling0423@gmail.com}   % email address
]{\inits{Y.L.L.}\fnm{Yan-Ling} \snm{Li}}
%\author[
%  addressref={aff1,aff2},
%  email={john.RS.Smith@cambridge.co.uk}
%]{\inits{J.R.S.}\fnm{John R.S.} \snm{Smith}}
%
%%%%%%%%%%%%%%%%%%%%%%%%%%%%%%%%%%%%%%%%%%%%%%
%%                                          %%
%% Enter the authors' addresses here        %%
%%                                          %%
%% Repeat \address commands as much as      %%
%% required.                                %%
%%                                          %%
%%%%%%%%%%%%%%%%%%%%%%%%%%%%%%%%%%%%%%%%%%%%%%

\address[id=aff1]{%                           % unique id
  \orgdiv{College of Physics and Electronic Information},             % department, if any
  \orgname{Gannan Normal University},          % university, etc
 \postcode{341000},
  \city{Ganzhou},                              % city
  \cny{China}                                    % country
}
\address[id=aff2]{%
  \orgdiv{School of Information Engineering},
  \orgname{Jiangxi University of Science and Technology},
 \postcode{341000},
  \city{Ganzhou},                              % city
  \cny{China}                                    % country
}

%%%%%%%%%%%%%%%%%%%%%%%%%%%%%%%%%%%%%%%%%%%%%%
%%                                          %%
%% Enter short notes here                   %%
%%                                          %%
%% Short notes will be after addresses      %%
%% on first page.                           %%
%%                                          %%
%%%%%%%%%%%%%%%%%%%%%%%%%%%%%%%%%%%%%%%%%%%%%%

%\begin{artnotes}
%%\note{Sample of title note}     % note to the article
%\note[id=n1]{Equal contributor} % note, connected to author
%\end{artnotes}

\end{fmbox}% comment this for two column layout

%%%%%%%%%%%%%%%%%%%%%%%%%%%%%%%%%%%%%%%%%%%%%%%
%%                                           %%
%% The Abstract begins here                  %%
%%                                           %%
%% Please refer to the Instructions for      %%
%% authors on https://www.biomedcentral.com/ %%
%% and include the section headings          %%
%% accordingly for your article type.        %%
%%                                           %%
%%%%%%%%%%%%%%%%%%%%%%%%%%%%%%%%%%%%%%%%%%%%%%%

\begin{abstract} % abstract

High-dimensional teleportation provides various benefits in quantum networks and repeaters, but all these advantages rely on the high-quality distribution of high-dimensional entanglement over a noisy channel. It is essential to consider correlation effects when two entangled qutrits travel sequentially through the same channel. 
In this paper, we present two strategies for enhancing qutrit teleportation in correlated amplitude damping (CAD) noise by weak measurement (WM) and environment-assisted measurement (EAM). The fidelity of both approaches has been dramatically improved due to the probabilistic nature of WM and EAM. We have observed that the correlation effects of CAD noise result in an increase in the probability of success. A comparison has demonstrated that the EAM scheme usually outperforms the WM scheme in regard to fidelity. Our research expands the capabilities of WM and EAM as quantum techniques to combat CAD noise in qutrit teleportation, facilitating the development of advanced quantum technologies in high-dimensional systems.
%\parttitle{First part title} %if any
%Text for this section.

%\parttitle{Second part title} %if any
%Text for this section.
\end{abstract}

%%%%%%%%%%%%%%%%%%%%%%%%%%%%%%%%%%%%%%%%%%%%%%
%%                                          %%
%% The keywords begin here                  %%
%%                                          %%
%% Put each keyword in separate \kwd{}.     %%
%%                                          %%
%%%%%%%%%%%%%%%%%%%%%%%%%%%%%%%%%%%%%%%%%%%%%%

\begin{keyword}
\kwd{Quantum teleportation}
\kwd{Correlated amplitude damping noise}
\kwd{Weak measurement}
\kwd{Environment-assisted measurement}
\end{keyword}

% MSC classifications codes, if any
%\begin{keyword}[class=AMS]
%\kwd[Primary ]{}
%\kwd{}
%\kwd[; secondary ]{}
%\end{keyword}

%
%\end{fmbox}% uncomment this for two column layout

\end{frontmatter}

%%%%%%%%%%%%%%%%%%%%%%%%%%%%%%%%%%%%%%%%%%%%%%%%
%%                                            %%
%% The Main Body begins here                  %%
%%                                            %%
%% Please refer to the instructions for       %%
%% authors on:                                %%
%% https://www.biomedcentral.com/getpublished %%
%% and include the section headings           %%
%% accordingly for your article type.         %%
%%                                            %%
%% See the Results and Discussion section     %%
%% for details on how to create sub-sections  %%
%%                                            %%
%% use \cite{...} to cite references          %%
%%  \cite{koon} and                           %%
%%  \cite{oreg,khar,zvai,xjon,schn,pond}      %%
%%                                            %%
%%%%%%%%%%%%%%%%%%%%%%%%%%%%%%%%%%%%%%%%%%%%%%%%

%%%%%%%%%%%%%%%%%%%%%%%%% start of article main body
% <put your article body there>

%%%%%%%%%%%%%%%%
%% Background %%
%%
\section*{1 Introduction}

\label{intro}
Quantum teleportation is not only a fascinating protocol in the theory of quantum information \cite{Bennett1993},
but also a fundamental part of long-distance quantum communication and quantum networks \cite{Nielsen2000} since it allows the non-local transmission of an unknown quantum state. Over the last thirty years, outstanding progress has been achieved in both the theoretical and experimental study of quantum teleportation \cite{Bouwmeester1997,Furusawa1998,Braunstein1998,Kim2001,Riebe2004,Barrett2004,Olms2009,Pfaff2014,Wang2015,Stefano2015}. In particular, there has been very rapid development in long-distance quantum teleportation, such as photonic quantum teleportation based on optical fibre networks and free-space satellites \cite{Sun2016,Valivarthi2016,Valivarthi2020,Ren2017,Lu2022,Raya2023}.

While the qubit (two-level system) serves as the basic unit of quantum information, a single particle in the actual world commonly has numerous degrees of freedom, thereby comprising a high-dimensional quantum system. Therefore, it is also of substantial interest to understand the functions of higher-level systems, such as the qutrit (three-level system) or the qudit (d-level system). In fact, quantum information protocols that rely on qutrits or qudits possess certain advantages over qubit-based schemes due to the additional dimensions in Hilbert space \cite{Lanyon2009}. These benefits encompass, but are not confined to an enhanced channel capacity \cite{Barreiro2008, Pirandola2017, Miller2019}, improved noise resilience \cite{Cerf2002, Ecker2019}, more efficient encoding of quantum information \cite{Cover2012}, a more reliable simulation of quantum dynamics \cite{Neeley2009,Blok2021}, and an increased sensitivity in quantum imaging schemes \cite{Fickler2012}. These advantages are particularly intriguing in the upcoming noisy intermediate-scale quantum (NISQ) era since the ``quality'' is of critical importance \cite{Preskill2018}. Recently, there has been significant progress in high-dimensional quantum teleportation experiments \cite{Luo2019,Hu2020b,Qiu2021,Sephton2021,Hu2023}.

%It is well known that the quantum superiority of quantum teleportation is mainly due to the initial shared entanglement between the sender and the receiver. 

The initial shared entanglement between the sender and the receiver is widely acknowledged as the primary reason for the superiority of quantum teleportation. However, quantum entanglement is very fragile in real-world environments. Therefore, the noise in the channel is a significant constraint on the fidelity of quantum teleportation. In the last decades, more attention has been paid to dealing with the noise in quantum teleportation \cite{Knoll2014,Fortes2015,Fonseca2019}. However, the majority of previous investigations have concentrated on the memoryless channel, where the noises are assumed to have different origins and are treated independently. While this presumption is reasonable in specific physical scenarios, it may not be justifiable in numerous more practical circumstances \cite{Macchiavello2002,Banaszek2004,Arrigo2007,Plenio2007,Arrigo2013}. The noises noises may share a common origin, resulting in their correlation with one another \cite{Lupke2020,Wilen2021}. For example, successive use of the same channel may cause the correlated noise between inputs, as the channel may retain memory between successive transmissions. 
Unlike the uncorrelated channel, which can be expressed as a tensor product of independent and identical completely positive, trace-preserving (CPTP) maps, the Kraus operators of the correlated channel map cannot be expressed in a tensor product form. The study of the correlated channel has attracted considerable attention in the field of quantum information, since memory effects cannot be avoided during high rate transmission \cite{Arrigo2015,Xiao2016a,Xu2019,Jeong2019,Seida2021,Sk2022,Sun2022,Li2022,Li2023,Lan2023}. 

\begin{figure}
 \includegraphics[width=0.8\textwidth]{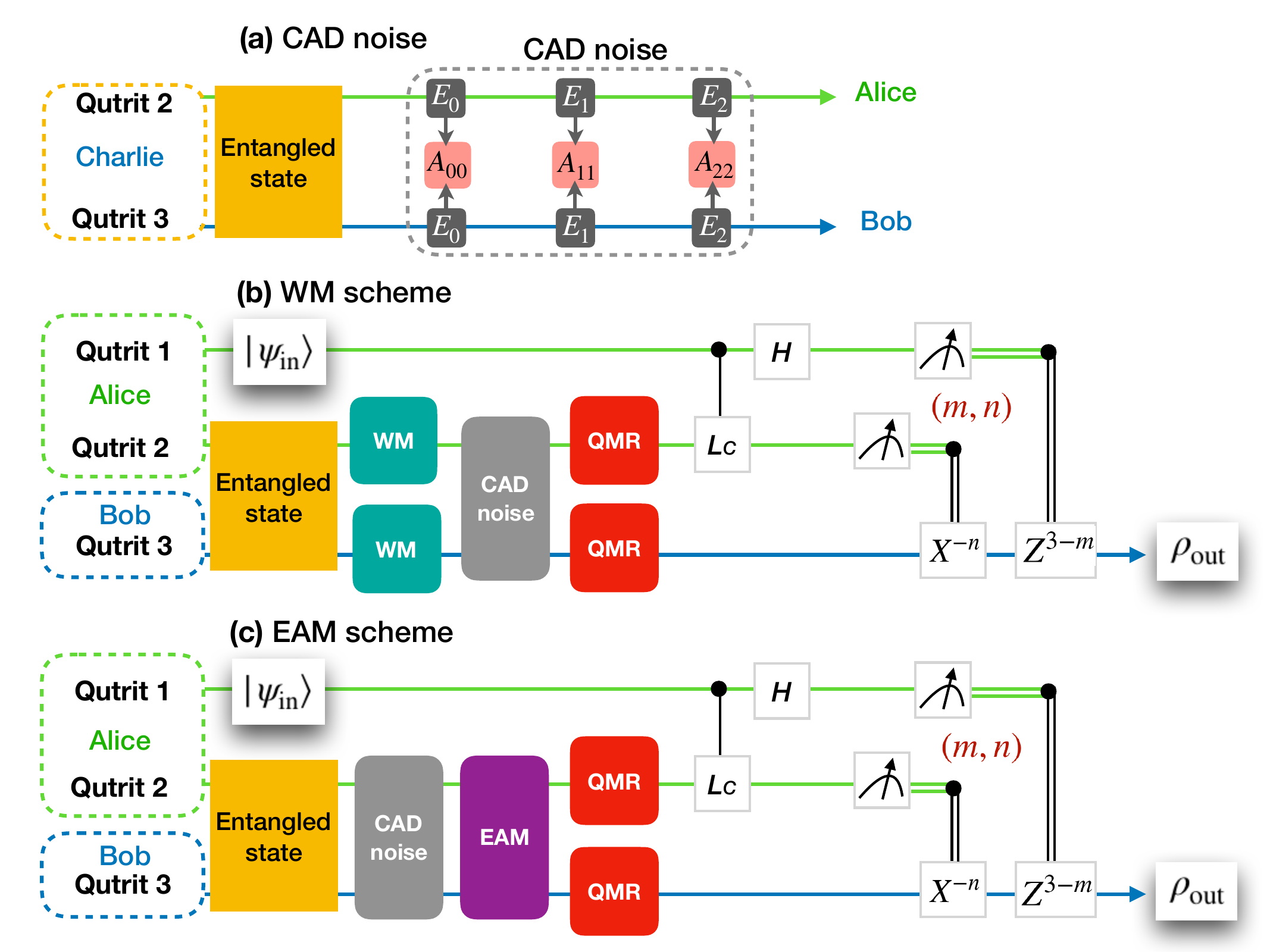}
\caption{(color online) (a) The diagram of CAD noise during the distribution of entangled state. Each qutrit will pass through the public channel with CAD noise. $E_{i}$, ($i = 0,1,2$) are the Kraus operators of AD noise, while the operators $A_{kk}$ with ($kk=00,11,22$) represent the correlation of noise between two uses of the lossy channel which originates from the much higher rate of consecutive transmission making the local environment retains a memory.
 (b) The circuit of qutrit teleportation with the assistance of WM and QMR.  (c) The circuit of qutrit teleportation with the assistance of EAM and QMR. $H$, $L_{C}$, $X$ and $Z$ are generalized logic gates for qutrit. $(m,n)$ denotes the result of the joint measurement performed by Alice.}
\label{Fig1}       % Give a unique
\end{figure}

In this paper, we present two techniques for improving the fidelity of qutrit teleportation in the presence of correlated amplitude damping (CAD) noise, through the use of weak measurement (WM) and environment-assisted measurement (EAM), as displayed in Fig.~\ref{Fig1}. 
Within the WM approach, pre-WM and post quantum measurement reversal (QMR) are performed on the system once before and once after it passes through the CAD channel. The purpose of pre-WM is to bring the initially entangled state closer to a `lethargic' state that is more resilient to the CAD noise, while the post-QMR aims to eliminate the effects of the pre-WM and the CAD noise, thus restoring the teleported state. The power of WM to combat AD noise has been demonstrated in many theoretical and experimental works \cite{Korotkov2006,Katz2008,Korotkov2010,Kim2012,Man2012,Wang2014,Zhang2015,Xiao2016b,Xiao2018,Xiao2020,Li2021,Harraz2022,Rodriguez2023}. Nevertheless, the potential of WM to handle CAD noise, particularly qutrit CAD noise, has yet to be explored. Although the fact that correlation effects in CAD noise can improve the fidelity of teleportation has been reported in Ref. \cite{Xu2022}, we emphasize that the combination of WM and QMR makes further efforts to improve fidelity.

In the EAM approach, a measurement such as photon counting is performed on the noisy environment coupled to the qutrit, followed by a QMR operation on the system conditioned by the measurement results. We find that the aid of 
EAM enables the fidelity of the teleportation to be almost completely recovered to 1, thereby eliminating the decoherence effect of the CAD noise. However, it should be acknowledged that the implementation of our approaches is subject to a certain probability. Fortunately, the correlation effects of CAD noise can increase the probability of success. It is interesting to note that the EAM scheme is mostly superior to WM scheme both in terms of fidelity and probability of success. The reason for this is that the post-performed EAM gathers information from both the system and the noisy channel.

The structure of this paper is as follows. In Sec. 2, we introduce some related concepts such as quantum teleportation, CAD noise, WM and QMR for the qutrit case. In Sec. 3, we then show how the fidelity of teleportation could be enhanced by WM and QMR. In Sec. 4, we propose another scheme to improve the fidelity with the assistance of EAM and QMR. A comparison of the two schemes is discussed in detail in Sec.5. In Sec.6, we extend the results to the case where the private channels are also affected by AD noise. Finally, the conclusions are summarized in Sec. 7.

\section*{2 Basic concepts}
 
\label{sec2}
\section*{2.1 CAD noise for qutrits}
\label{sec2.1}
Amplitude damping noise is a prototype model of a dissipative interaction between a quantum system and its environment, such as spontaneous emission in the atomic system or photon loss in the optical system \cite{Nielsen2000}. Assuming that two qutrits successively pass through an AD channel, the correlation effects should be considered if the relaxation time of the channel is not too short compared with the time interval between the two consecutive qutrits entering the channel. In this case, the channel is a CAD channel which cannot be factorized as those on each one, as shown in Fig.~\ref{Fig1}(a). 

As we all know, an arbitrary completely positive, trace-preserving (CPTP) map $\mathcal{E}$ can be written as $\mathcal{E}(\rho)=\sum_{j}E_{j}\rho_{0} E_{j}^{\dagger}$ in the operator-sum representation, where $E_{j}$ are the Kraus operators. Note that for the uncorrelated AD channel, the overall dynamical map can be expressed as a tensor product of the individual AD maps: $\mathcal{E}_{\rm AD}^{(n)}(\rho)=\mathcal{E}_{\rm AD}^{\otimes n}(\rho_{0})$. However, such a tensorial decomposition is not valid for the CAD channel. The map of a CAD channel is a combination of the uncorrelated AD channel and fully correlated AD (FCAD) channel:
\begin{eqnarray}
\label{eq1}
\mathcal{E}_{\rm CAD}(\rho)&=&(1-\mu)\mathcal{E}_{\rm AD}^{\otimes2}(\rho)+\mu\mathcal{E}_{\rm FCAD}(\rho),\\
&=&(1-\mu)\sum_{i,j=0}^{2}E_{ij}\rho E_{ij}^{\dagger}+\mu\sum_{k=0}^{2}A_{kk}\rho A_{kk}^{\dagger},\nonumber
\end{eqnarray}
where $\mu\in[0,1]$ is the correlation parameter. We can recover the uncorrelated AD channel by setting $\mu=0$ and
obtain the FCAD channel if $\mu=1$. The explicit expressions of the Kraus operators
$E_{ij}=E_{i}\otimes E_{j}$ and $A_{kk}$ are determined by solving the Lindblad equation (please see the details in Appendix \ref{app1}) \cite{Jeong2019,Xu2022} 
\begin{eqnarray}
\label{eq2}
E_{0}=\left(\begin{array}{ccc}1 & 0 & 0 \\0 & \sqrt{\overline{d}_1} & 0 \\0 & 0 & \sqrt{\overline{d}_2}\end{array}\right),E_{1}=\left(\begin{array}{ccc}0 & \sqrt{d_1} & 0 \\0 & 0 & 0 \\0 & 0 & 0\end{array}\right),E_{2}=\left(\begin{array}{ccc}0 & 0 & \sqrt{d_2} \\0 & 0 & 0 \\0 & 0 & 0\end{array}\right),
\end{eqnarray}
with $d_1=1-\exp(-\gamma_{10}t)$, $d_2=1-\exp(-\gamma_{20}t)$ and the symbol $\overline{d}=1-d$. 
The Kraus operators of the correlated part are given as
\begin{eqnarray}
\label{eq3}
A_{00}&=&\left(\begin{array}{cccc}{\rm \bf I}_{4\times4} &  &  &  \\  & \sqrt{\overline{d}_1} &  &  \\  &   & {\rm \bf I}_{3\times3} &   \\  &   &   & \sqrt{\overline{d}_2}\end{array}\right),
A_{11}=\left(\begin{array}{c|c|c} {\rm \bf 0}_{1\times4}  & \sqrt{d_1} & {\rm \bf 0}_{1\times4}   \\
\hline
 {\rm \bf 0}_{4\times4}   & {\rm \bf 0}_{4\times1}   &  {\rm \bf 0}_{4\times4} \\
\hline
  {\rm \bf 0}_{4\times4} & {\rm \bf 0}_{4\times1}   & {\rm \bf 0}_{4\times4}  \end{array}\right),\nonumber\\
A_{22}&=&\left(\begin{array}{c|c|c} {\rm \bf O}_{1\times4}  &  {\rm \bf 0}_{1\times4}  & \sqrt{d_2}   \\
\hline
 {\rm \bf 0}_{4\times4}   & {\rm \bf 0}_{4\times4}   &  {\rm \bf 0}_{4\times1} \\
\hline
  {\rm \bf 0}_{4\times4} & {\rm \bf 0}_{4\times4}   & {\rm \bf 0}_{4\times1}  \end{array}\right),
\end{eqnarray}
where ${\rm \bf I}_{n\times n}$ is the $n$-dimensional identity matrix and $ {\rm \bf 0}_{n\times m}$ is the zero matrix of $n\times m$.

\section*{2.2 Standard quantum teleportation for qutrit}
\label{sec2.2}
Quantum teleportation aims to transfer an unknown quantum state from the sender Alice to the receiver Bob. In what follows we discuss how to implement the teleportation of a qutrit state 
\begin{equation}
\label{eq4}
|\psi\rangle_{\rm in}=\alpha|0\rangle+\beta|1\rangle+\delta|2\rangle,
\end{equation}
with $\alpha=\cos\theta_{1}$, $\beta=\sin\theta_{1}\cos\theta_{2}e^{i \phi_{1}}$, $\delta=\sin\theta_{1}\sin\theta_{2}e^{i \phi_{2}}$ and ($0<\theta_{1},\theta_{2}\leq \frac{\pi}{2}$, $0<\phi_{1},\phi_{2}\leq 2\pi$).

(i) The first step in teleportation is to establish the entanglement between Alice and Bob. This could be done in a number of ways.
Here, we assume that the entangled particle pairs (qutrit 2 and qutrit 3) are generated by the third party Charlie
\begin{equation}
\label{eq5}
|\Phi\rangle_{(23)}=\frac{1}{\sqrt{3}}(|00\rangle+|11\rangle+|22\rangle),
\end{equation}
where the subscript ``(23)'' denotes the qutrits 2 and 3, respectively. 
Charlie distributes qutrit 2 and qutrit 3 to Alice and Bob through a public AD channel and then via two separate noiseless private channels. Because of the successive uses of the public AD channel, the correlation effects should be taken into account. Therefore, the noisy channel can be modelled as a CAD channel.

(ii) First, Alice interacts the qutrit 1 containing the teleported information with her half of the entangled pair (qutrit 2) thorugh the logic gates $L_{C}$ and $H$, where $H=\frac{1}{\sqrt{3}} \sum_{m, n=0}^{2} \mathrm{e}^{2 \pi \mathrm{i} m n / 3}|m\rangle\langle n|$ is the Hadamard gate of the qutrit and $L_{C}|m\rangle \otimes|n\rangle=|m\rangle \otimes|n \ominus m\rangle$ is the CNOT left shift gate for two qutrits, where $\ominus$ denotes subtraction module $3$. Then Alice performs measurements on her particles  (qutrit 1 and qutrit 2). Finally, Alice sends her result $(m,n)$ to Bob through a classical channel.

(iii) According to the result $(m,n)$, Bob applies a series of recovery operations $X^{-n}$ and $Z^{3-m}$ on his particle to obtain the teleported state. $X^{n}|k\rangle=|k\oplus n\rangle$ and $Z^{m}|k\rangle=\mathrm{e}^{2 \pi \mathrm{i} k m/ 3}|k\rangle$ are NOT gate and phase gate for single qutrit.

In the noiseless case, Bob can obtain the exact state of the form of Eq. (\ref{eq4}). However, the inevitable noise in the procedure will make the state obtained by Bob different from the teleported state, i.e., the fidelity is less than 1.

\section*{2.3 WM and QMR for qutrit}
\label{sec2.2}
The WM that we consider is a positive operator valued measure (POVM), which is different from the so-called strong measurement (i.e., von Neumann projective measurement). WM does not completely destroy the measured system, thereby retaining the measured state reversible. Although the WM for qubit has been widely explored in many previous literatures \cite{Korotkov2006,Katz2008,Korotkov2010,Kim2012,Man2012,Wang2014,Zhang2015,Xiao2016b,Xiao2018,Xiao2020,Li2021,Harraz2022,Rodriguez2023}, the specific form of WM for qutrit was first proposed by Xiao in Ref. \cite{Xiao2013,Xiao2014}. The WM for single-qutrit is written as
\begin{eqnarray}
\label{eq6}
\mathcal{M}=\left(\begin{array}{ccc}1 & 0 & 0 \\0 & \sqrt{\overline{p}_1} & 0 \\0 & 0 & \sqrt{\overline{p}_2}\end{array}\right).
\end{eqnarray}
The parameters $p_1,p_2\in[0,1]$ are usually known as the strengths of WM. $\mathcal{M}$ and the other two POVM elements $\mathcal{M}_{1}={\rm diag}(0,\sqrt{p_1},0)$ and $\mathcal{M}_{2}={\rm diag}(0,0,\sqrt{p_2})$ satisfy the completeness relation $\mathcal{M}\mathcal{M}^{\dagger}+\mathcal{M}_{1}\mathcal{M}_{1}^{\dagger}+\mathcal{M}_{2}\mathcal{M}_{2}^{\dagger}=I$. 
However, the measurement operators $\mathcal{M}_{1}$ and $\mathcal{M}_{2}$ are not reversible. We only focus on the measurement operator $\mathcal{M}$, which maps the qutrit to a state that can still be recovered by proper operations, e.g., QMR. 

Physically, the measurement operator $\mathcal{M}$ could be realized by adding a detector to monitor the dissipation of the system. Whenever there is a click, we discard the result. Therefore, this post-selection removes the outcomes of $\mathcal{M}_{1}$ and $\mathcal{M}_{2}$ and keeps the result of $\mathcal{M}$ (no click). WM can be implemented with different technologies on different platforms for the qubit system \cite{Katz2008,Kim2012}. For a $V$-type qutrit system, we can choose the energy structure of the $^{87}\rm{Rb}$ atom \cite{Lloyd2001}. The ground state $|0\rangle$ corresponds to the level $|F=1, m_{F}=0\rangle$ of $5^{2}S_{1/2}$, while the two excited states $|1\rangle$ and $|2\rangle$ correspond to the degenerate levels $|F=1, m_{F}=1\rangle$ and $|F=1, m_{F}=-1\rangle$ of $5^{2}P_{1/2}$. The transitions $|1\rangle\rightarrow|0\rangle$ and $|2\rangle\rightarrow|0\rangle$ emit right- and left-circularly polarized photons, respectively.
In the case that a photon is detected (regardless of right- or left-circularly polarized photon), the result should be discarded since it is the outcome of $\mathcal{M}_{1}$ or $\mathcal{M}_{2}$. Conversely, a no click result represents the implementation of $\mathcal{M}$. The parameters $p_1$ and $p_2$ can be determined by counting the right- and left-circularly polarized photons.

The QMR $\mathcal{R}$ for single-qutrit is 
\begin{eqnarray}
\label{eq7}
\mathcal{R}=\left(\begin{array}{ccc}\sqrt{\overline{q}_1\overline{q}_2} & 0 & 0 \\0 & \sqrt{\overline{q}_2} & 0 \\0 & 0 & \sqrt{\overline{q}_1}\end{array}\right)=\mathcal{F}\mathcal{M}(q_{1},q_{2})\mathcal{F}\mathcal{M}(q_{1},q_{2})\mathcal{F},
\end{eqnarray}
where $\mathcal{F}$ is the trit-flip operation $\mathcal{F}=|0\rangle\langle 2|+|1\rangle\langle 0|+|2\rangle\langle 1|$. $\mathcal{M}(q_{1},q_{2})$ has the same form as Eq. (\ref{eq6}) but with the parameters $q_1$ and $q_2$ that denote the strength of QMR. Although the QMR cannot be implemented as an evolution governed by a suitable Hamiltonian, it can nevertheless be achieved with a certain probability through the combination of the WM and trit-flip operations. The second equality of Eq.~(\ref{eq7}) suggests that the QMR can be realized by the following five sequential operations on the qutrit: trit-flip ($\mathcal{F}$), qutrit WM
($\mathcal{M}$), trit-flip ($\mathcal{F}$), qutrit WM ($\mathcal{M}$), and trit-flip ($\mathcal{F}$). The
trit-flip operation $\mathcal{F}$ can be realized by a $\pi$ pulse
applied on the transition $|1\rangle\leftrightarrow|2\rangle$ and
followed by another $\pi$ pulse to interchange the populations
between $|0\rangle$ and $|1\rangle$, i.e., by the series of two
$\pi$ pulses
$\pi^{|1\rangle\leftrightarrow|2\rangle}\pi^{|0\rangle\leftrightarrow|1\rangle}$ \cite{Scully1997}.

\section*{3 Enhancing the fidelity of teleportation by WM and QMR}
\label{sec3}
We are now in a position to demonstrate the basic idea of our WM scheme. In order to enhance the fidelity of teleportation, the pre-WMs and post-QMRs are carried out before and after the CAD noise, as illustrated in Fig.~\ref{Fig1}(b). 
Charlie performs WMs on the entangled pairs immediately after preparing the initial entangled state. These pre-WMs are
designed to project the entangled state (\ref{eq5}) onto a state that is insensitive to CAD noise.
Then he distributes them through the CAD channel. When Alice and Bob receive their qutrits, they perform QMRs on their own qutrit. The QMRs aim to recover the initial entanglement between qutrits 2 and 3. After these three steps, Alice and Bob will share the entangled state (unnormalized)

\begin{equation}
\rho_{(23)}^{\rm WM}= \mathcal{R}_{(23)}\mathcal{E}_{\rm CAD}\Big[\mathcal{M}_{(23)}\big(|\Phi\rangle_{(23)}\langle\Phi|\big)\mathcal{M}_{(23)}^{\dagger}\Big]\mathcal{R}_{(23)}^{\dagger}.
\label{eq8}
\end{equation}
$\mathcal{M}_{(23)}$ and $\mathcal{R}_{(23)}$ are defined as $\mathcal{M}_{(23)}=\mathcal{M}_{(2)}\otimes\mathcal{M}_{(3)}$
 and $\mathcal{R}_{(23)}=\mathcal{R}_{(2)}\otimes\mathcal{R}_{(3)}$. $\rho_{(23)}^{\rm WM}$ is a $9\times 9$ matrix which has following non-zero elements in the basis $\{|j,k\rangle=|3j+k+1\rangle\}$:
\begin{eqnarray}
\label{eq9}
\rho_{11}&=&\frac{1}{3}\{\overline{q}_{1}^{2}\overline{q}_{2}^{2}[1+(\overline{\mu}d_{1}^{2}+\mu d_{1})\overline{p}_{1}^{2}+(\overline{\mu}d_{2}^{2}+\mu d_{2})\overline{p}_{2}^{2}]\}, \nonumber\\
\rho_{22}&=&\rho_{44}= \frac{1}{3}\overline{\mu}d_{1}\overline{d}_{1}\overline{p}_{1}^{2}\overline{q}_{1}\overline{q}_{2}^{2},\nonumber\\
\rho_{33}&=&\rho_{77}= \frac{1}{3}\overline{\mu}d_{2}\overline{d}_{2}\overline{p}_{2}^{2}\overline{q}_{1}^{2}\overline{q}_{2},\nonumber\\
\rho_{55}&=&\frac{1}{3}(\overline{\mu}\overline{d}_{1}^{2}+\mu \overline{d}_{1})\overline{p}_{1}^{2}\overline{q}_{2}^{2},\nonumber\\
\rho_{99}&=&\frac{1}{3}(\overline{\mu}\overline{d}_{2}^{2}+\mu \overline{d}_{2})\overline{p}_{2}^{2}\overline{q}_{1}^{2},\\
\rho_{15}&=&\rho_{51}^{*}= \frac{1}{3}(\overline{\mu}\overline{d}_{1}+\mu \sqrt{\overline{d}_{1}})\overline{p}_{1}\overline{q}_{1}\overline{q}_{2}^{2},\nonumber\\
\rho_{19}&=&\rho_{91}^{*}= \frac{1}{3}(\overline{\mu}\overline{d}_{2}+\mu \sqrt{\overline{d}_{2}})\overline{p}_{2}\overline{q}_{1}^{2}\overline{q}_{2},\nonumber\\
\rho_{59}&=&\rho_{95}^{*}= \frac{1}{3}(\overline{\mu}\overline{d}_{1}\overline{d}_{2}+\mu \sqrt{\overline{d}_{1}\overline{d}_{2}})\overline{p}_{1}\overline{p}_{2}\overline{q}_{1}\overline{q}_{2}.\nonumber
\end{eqnarray}

Through the procedure of quantum teleportation, Bob obtains the state
\begin{eqnarray}
\label{eq10}
\rho_{\rm out}^{\rm WM} &=& \frac{1}{N}\left[\begin{array}{ccc} \epsilon_{00} & \epsilon_{01} & \epsilon_{02} \\ \epsilon_{10} & \epsilon_{11} & \epsilon_{12}\\ \epsilon_{20} &\epsilon_{21} & \epsilon_{22} \end{array}\right],
\end{eqnarray}
where $\epsilon_{00}=(\rho_{11}+\rho_{55}+\rho_{99})\alpha^{2}+(\rho_{33}+\rho_{44})|\beta|^{2}+(\rho_{22}+\rho_{77})|\delta|^{2}$, $\epsilon_{11}=(\rho_{22}+\rho_{77})\alpha^{2}+(\rho_{11}+\rho_{55}+\rho_{99})|\beta|^{2}+(\rho_{33}+\rho_{44})|\delta|^{2}$,
$\epsilon_{22}=(\rho_{33}+\rho_{44})\alpha^{2}+(\rho_{22}+\rho_{77})|\beta|^{2}+(\rho_{11}+\rho_{55}+\rho_{99})|\delta|^{2}$,
$\epsilon_{01}=\epsilon_{10}^{*}= \alpha \beta^{*}(\rho_{15}+\rho_{59}+\rho_{91})$,
$\epsilon_{02}=\epsilon_{20}^{*}= \alpha \delta^{*}(\rho_{19}+\rho_{51}+\rho_{95})$,
$\epsilon_{12}=\epsilon_{21}^{*}= \beta \delta^{*}(\rho_{15}+\rho_{59}+\rho_{91})$,
and $N=\rho_{11}+\rho_{22}+\rho_{33}+\rho_{44}+\rho_{55}+\rho_{77}+\rho_{99}$ is the normalization factor.
%\begin{eqnarray}
%\label{eq11}
%\epsilon_{00}=&&(\rho_{11}+\rho_{55}+\rho_{99})\alpha^{2}+(\rho_{33}+\rho_{44})|\beta|^{2}+(\rho_{22}+\rho_{77})|\delta|^{2}, \nonumber\\
%\epsilon_{11}=&&(\rho_{22}+\rho_{77})\alpha^{2}+(\rho_{11}+\rho_{55}+\rho_{99})|\beta|^{2}+(\rho_{33}+\rho_{44})|\delta|^{2},\nonumber\\
%\epsilon_{22}=&&(\rho_{33}+\rho_{44})\alpha^{2}+(\rho_{22}+\rho_{77})|\beta|^{2}+(\rho_{11}+\rho_{55}+\rho_{99})|\delta|^{2},\nonumber\\
%\epsilon_{01}=&&\epsilon_{10}^{*}= \alpha \beta^{*}(\rho_{15}+\rho_{59}+\rho_{91}),\\
%\epsilon_{02}=&&\epsilon_{20}^{*}= \alpha \delta^{*}(\rho_{19}+\rho_{51}+\rho_{95}),\nonumber\\
%\epsilon_{12}=&&\epsilon_{21}^{*}= \beta \delta^{*}(\rho_{15}+\rho_{59}+\rho_{91}),\nonumber
%\end{eqnarray}

The fidelity of this teleportation is now given by
\begin{eqnarray}
\label{eq12}
F^{\rm WM}&&= _{\rm in}\langle\psi|\rho_{\rm out}^{\rm WM}|\psi\rangle_{\rm in}\\
&&=A_{1}+(\alpha^{2}|\beta|^{2}+\alpha^{2}|\delta|^{2}+|\beta|^{2}|\delta|^{2})(A_{2}+A_{3}-2A_{1}),\nonumber
\end{eqnarray}
where $A_{1}=(\rho_{11}+\rho_{55}+\rho_{99})/N$, $A_{2}=(\rho_{22}+\rho_{33}+\rho_{44}+\rho_{77})/N$ and $A_{3}=(\rho_{15}+\rho_{51}+\rho_{19}+\rho_{91}+\rho_{59}+\rho_{95})/N$.

Considering that the teleported state $|\psi\rangle_{\rm in}$ is unknown, one should calculate the average fidelity. Since Eq. (\ref{eq12}) only contains the terms $|\beta|^2$ and $|\delta|^2$, hence the average fidelity could be written as 
\begin{eqnarray}
\label{eq13}
\langle F\rangle^{\rm WM}&&=\frac{2}{\pi^{2}}\int_{0}^{\frac{\pi}{2}}\int_{0}^{\frac{\pi}{2}}\int_{0}^{2\pi}\int_{0}^{2\pi}F^{\rm WM}\sin^{3}\theta_{1}\cos\theta_{1}d\theta_{1}d\phi_{1}\sin\theta_{2}\cos\theta_{2}d\theta_{2}d\phi_{2},\nonumber\\
&&=\frac{1}{4}(1+A_{1}+A_{3}).
\end{eqnarray}
It is obvious that the fidelity $\langle F\rangle^{\rm WM}$ is highly dependent on the strength of the WM and QMR. In oder to fight the CAD noise and obtain the maximum average fidelity, it is crucial to choose the optimal strength of the post-QMRs. 

It is generally believed that entanglement is the key resource of quantum teleportation and that entanglement degradation is directly responsible for the fidelity loss. Therefore, the most intuitive way to improve the average fidelity is to choose a post-QMR with an appropriate strength that keeps the entangled state $\rho_{(23)}^{\rm WM}$ as close as possible to the initial state $|\Phi\rangle_{(23)}$.
The decays of both uncorrelated the AD noise and the FCAD noise could be regarded as quantum transitions that lead to the excited states jumping to the only ground state with certain probabilities. 
Thus, one can technically use the trick of `unraveling' the excitation into `jump' and `no-jump' scenarios and work with pure states \cite{Scully1997}. It can be shown that in the CAD channel the optimal strength of the QMRs yield to (please see the details in Appendix \ref{app2}) 
\begin{eqnarray}
\label{eqb14}
\overline{q}_{\rm 1,opt}^{\rm WM}&=&\overline{p}_{1}\Big(\overline{\mu}\overline{d}_{1}+\mu \sqrt{\overline{d}_{1}}\Big),\\
\overline{q}_{\rm 2,opt}^{\rm WM}&=&\overline{p}_{2}\Big(\overline{\mu}\overline{d}_{2}+\mu \sqrt{\overline{d}_{2}}\Big).\nonumber
\end{eqnarray}

For the sake of simplicity, we assume $d_1=d_2=d$, $p_1=p_2=p$ and consequently $q_1=q_2=q$. Thus the optimal strength of post-QMR is
\begin{equation}
\label{eq14}
q_{\rm opt}^{\rm WM}=1-\overline{p}\Big(\overline{\mu}\overline{d}+\mu \sqrt{\overline{d}}\Big).
\end{equation}

%In the following two subsection, we propose two approaches to choose the strength of post-QMR. Both of them effectively enhance the average fidelity of teleportation. We name them as physical approach and mathematical approach. 
%For the sake of simplicity, we assume 
%$d_1=d_2=d$, $p_1=p_2=p$ to obtain the general analytic expressions are too complicated to present.

The average fidelity of Eq. (\ref{eq13}) will be written as 
\begin{eqnarray}
\label{eq15}
\langle F\rangle^{\rm WM}_{\rm opt}=\frac{1}{4}+\frac{\big[2(\overline{\mu}d+\mu)d \overline{p}^{2}+5\big](\overline{\mu}\sqrt{\overline{d}}+\mu )^{2} +4(\overline{\mu}\overline{d}+\mu )}{\big[8(\overline{\mu}d+\mu) d\overline{p}^{2}+4\big](\overline{\mu}\sqrt{\overline{d}}+\mu )^{2} +8(\overline{\mu}\overline{d}+\mu)+16 \overline{\mu} d (\overline{\mu}\overline{d}+\mu \sqrt{\overline{d}})}.\nonumber\\
\end{eqnarray}

Since the WMs and QMRs are not unitary, the improvement of fidelity is not deterministic but probabilistic. Under the condition of Eq. (\ref{eq14}), the probability of success is
\begin{equation}
\label{eq16}
P_{\rm opt}^{\rm WM}=\frac{1}{3}\overline{p}^{4}\overline{d}^{2}(\overline{\mu}\sqrt{\overline{d}}+\mu)^{2}\Big\{ \big[2(\overline{\mu}d+\mu)d \overline{p}^{2}+1\big](\overline{\mu}\sqrt{\overline{d}}+\mu)^{2} +2(\overline{\mu}\overline{d}+\mu )+4 \overline{\mu} d (\overline{\mu}\overline{d}+\mu \sqrt{\overline{d}}) \Big\}.
\end{equation}

\begin{figure*}
 \includegraphics[width=0.8\textwidth]{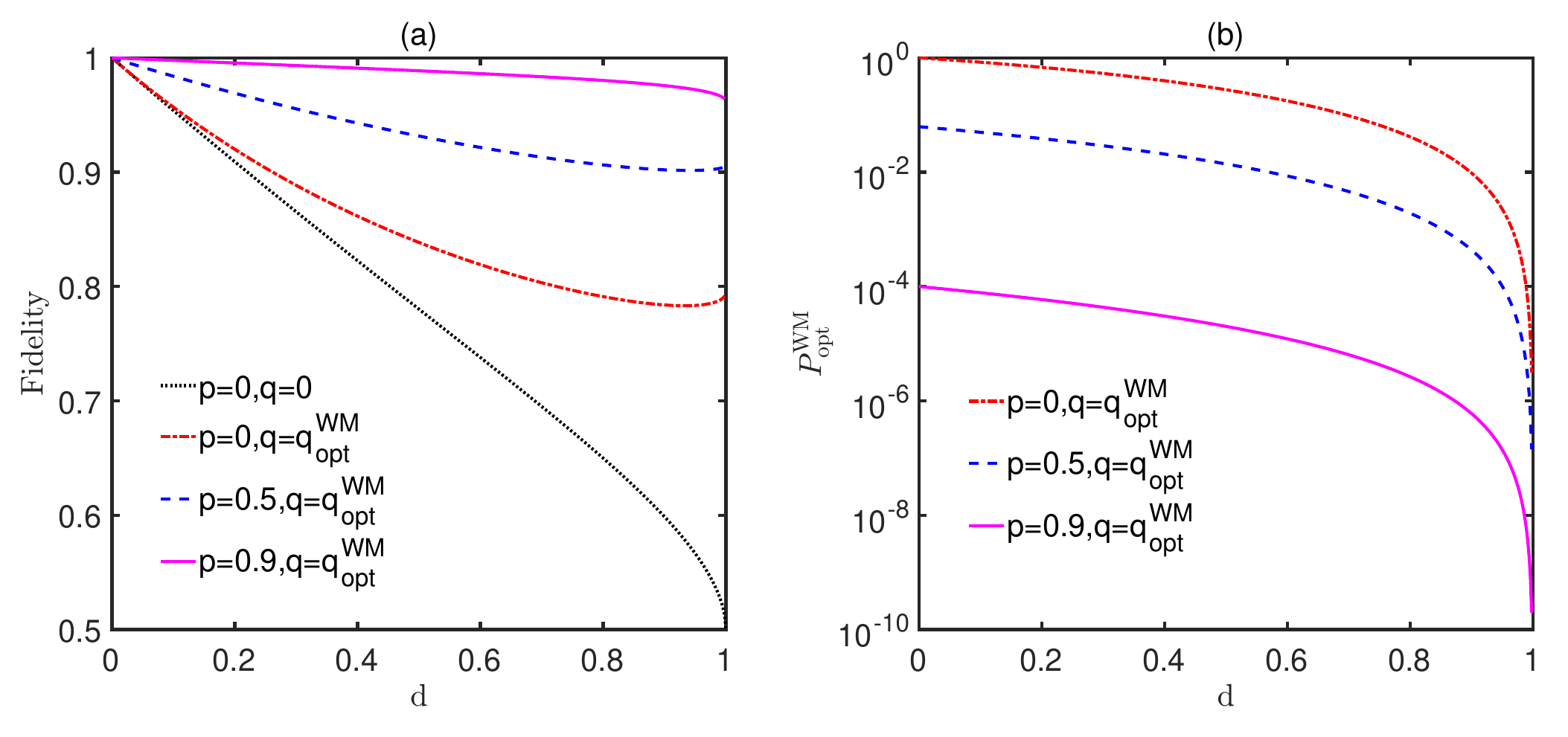}
\caption{(color online) (a) The average fidelity $\langle F\rangle_{\rm opt}^{\rm WM}$ and (b) the probability of success $P_{\rm opt}^{\rm WM}$ as a function of the noise strength $d$ for different values of $p$ with $\mu=0.8$.}
\label{Fig2}       % Give a unique
\end{figure*}

In Fig. \ref{Fig2}, we plot the average fidelity $\langle F\rangle_{\rm opt}^{\rm WM}$ and the probability of success $P_{\rm opt}^{\rm WM}$ as a function of the noise strength $d$ for different values of $p$. From Fig. \ref{Fig2}, we can summarize the following conclusions. 
(i) In the absence of WM and QMR, i.e., $p=q=0$, the average fidelity decreases as the noise strength increases, which directly indicates the degradation of the entanglement initially shared between Alice and Bob, since the fidelity depends on the entanglement. Note that the fidelity bound of classical teleportation in a 3-dimensional single-qutrit case is 1/2 \cite{Hayashi2005}. The quantum advantage is always preserved in the CAD noise, meaning that quantum teleportation is more robust to noise in higher-dimensional systems. 
(ii) The fidelity can be partially improved by the post-QMR even without the pre-WM ($p=0$). It could be considered as a simplified version of error correction based on prior knowledge of the CAD noise. 
(iii) The combination of pre-WM and post-QMR is indeed able to further enhance the fidelity of teleportation in the CAD noise. 
(iv) The probability of success decreases with increasing strength of pre-WM, which means that high fidelity teleportation is realized at the expense of low probability of success, as shown in Fig.~\ref{Fig2}(b).

\begin{figure*}
 \includegraphics[width=0.8\textwidth]{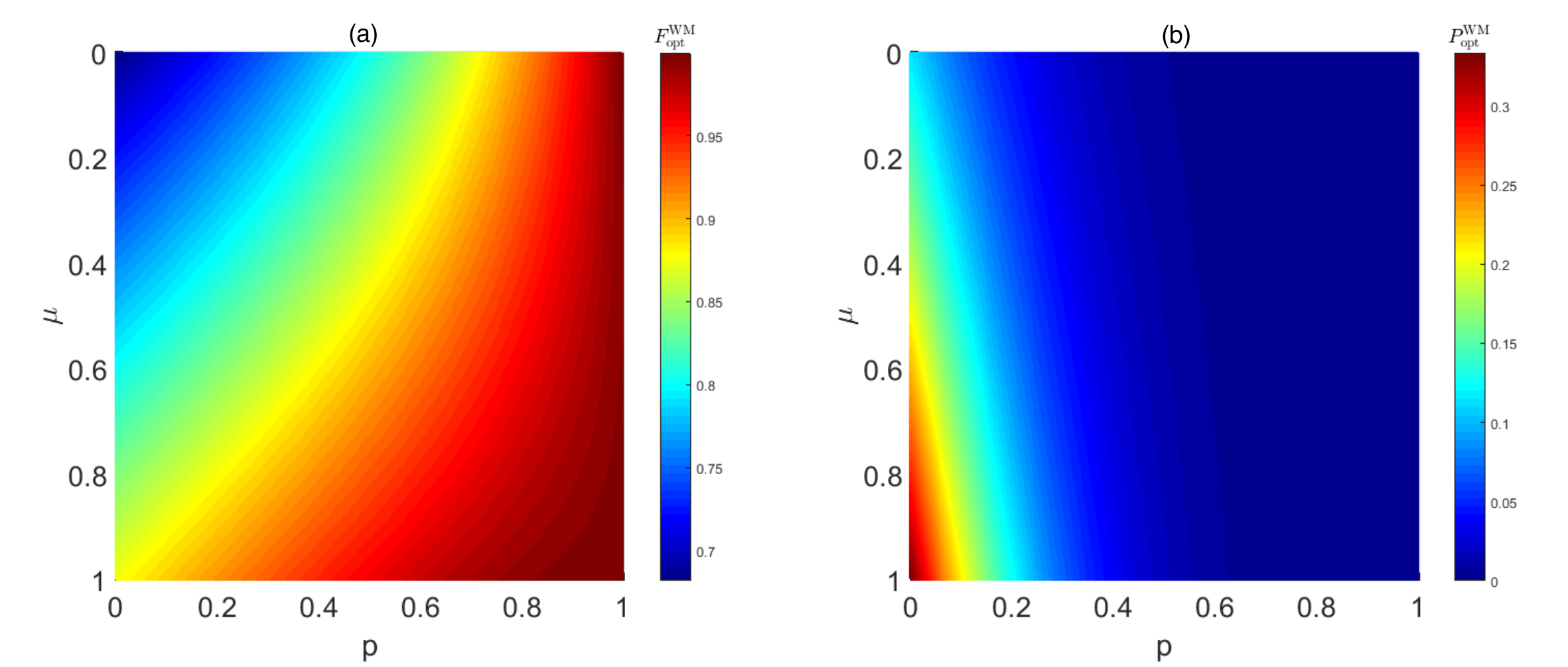}
\caption{(color online) (a) The average fidelity $\langle F\rangle_{\rm opt}^{\rm WM}$ and (b) the probability of success $P_{\rm opt}^{\rm WM}$ as a function of the correlated factor $\mu$ and WM strength $p$ with $d=0.6$.}
\label{Fig3}       % Give a unique
\end{figure*}

Figure \ref{Fig3}(a) displays the behavior of the fidelity $\langle F\rangle_{\rm opt}^{\rm WM}$ as a function of $\mu$ and $p$ at a fixed value of the noisy strength $d=0.6$. It is clear that the correlation effects have a positive effect on the behavior of the fidelity, which is in consistent with that obtained in Ref.~\cite{Xu2022}. On the other hand, the combined action of pre-WM and post-QMR can further enhance fidelity based on correlation effects. The higher the strength of the WM, the higher the fidelity. In the limit $p\rightarrow 1$, the fidelity is close to 1 because the entangled state $|\Phi\rangle_{(23)}$ is completely recovered by the optimal QMR. Although the optimal condition in Eq. (\ref{eq14}) requires prior knowledge of the noise, it can almost
completely eliminate the effect of the CAD noise without knowing the information of the teleported state.
However, it should be noted that the higher fidelity comes at the cost of a lower probability of success. Generally speaking, there is a trade-oﬀ between fidelity improvement and probability of success in these probabilistic schemes, and it should be carefully balanced in realistic situations. For some tasks it may be preferable to have as high a fidelity as possible regardless of the low probability of success, whereas for other tasks the probability of success is more important if the fidelity is high enough. Fortunately, the correlation effects can increase the probability of success to some extent, as shown in Fig. \ref{Fig3}(b).

\section*{4 Enhancing the fidelity of teleportation by EAM and QMR}
\label{sec4}
Unlike WM, which is applied to the quantum system before it enters into the noisy channel, EAM is acted on the environment when the quantum system passes through it \cite{Zhao2013}. 
Assuming that the environment is in a vacuum state $|0\rangle_{\rm env}$, then the evolution of the total system (i.e., qutrit plus environment) is determined by $\rho_{\rm tot}=U(\rho_{\rm s}(0)\otimes|0\rangle_{\rm env}\langle 0|)U^{\dagger}$, where $U=\exp(-i\hat{H}_{\rm tot}t)$ is the time-dependent evolution operator and $\rho_{\rm s}(0)$ is
the initial state of the system. The reduced density matrix of the system $\rho_{\rm s}(t)$ can be obtained by tracing over the environmental degrees of freedom, which yields to
\begin{eqnarray}
\label{eq18}
\rho_{\rm s}(t)&=&\sum_{n}{_{\rm env}}\langle n|U|0\rangle_{\rm env}\rho_{\rm s}(0)_{\rm env}\langle 0|U^{\dagger}|n\rangle_{\rm env}\nonumber\\
&=&\sum_{n}E_{n}\rho_{\rm s}(0)E_{n}^{\dagger},
\end{eqnarray}
where $E_{n}={_{\rm env}}\langle n|U|0\rangle_{\rm env}$ is the Kraus operator of the environmental noise. $\{|n\rangle_{\rm env}\}$ are the complete basis of the environment.

A photon counting measurement on the excitation of the environment will collapse the state of the environment into the $k^{\rm th}$ eigenstate if $k$ photons are collected. Subsequently, the system will be projected into a state that corresponds to the $k^{\rm th}$ outcome \cite{Zhao2013,WangZhao2014}: 
\begin{equation}
\label{eq19}
\rho^{k}_{\rm s}=E_{k}\rho_{\rm s}(0)E_{k}^{\dagger}.
\end{equation}
If the operator $E_{k}$ is reversible, then one can perform the reversal operations to recover the initial state of the system. Obviously, only $E_{00}$ and $A_{00}$ in Eq. (\ref{eq1}) satisfy the reversibility condition. Thus, we only focus on the measurement results of $k=0$ since they are reversible, while discarding the results of $k\neq0$. This post-selection allows the system to collapse into a specific state that can be recovered by QMR.

According to the procedure illustrated in Fig. \ref{Fig1}(c), after the sequences of CAD, EAM and QMRs, Alice and Bob will share the entangled state (unnormalized)
\begin{equation}
\label{eq20}
\rho_{(23)}^{\rm EAM}= \mathcal{R}_{(23)}\Big[(1-\mu)E_{00}|\Phi\rangle_{(23)}\langle\Phi|E_{00}^{\dagger}+\mu A_{00}|\Phi\rangle_{(23)}\langle\Phi|A_{00}^{\dagger}\Big]\mathcal{R}_{(23)}^{\dagger},
\end{equation}
which has the following non-zero elements:
\begin{eqnarray}
\label{eq21}
\rho'_{11}=&&\frac{1}{3}\{\overline{q}_{1}^{2}\overline{q}_{2}^{2}(\mu M_{1}+\overline{\mu}M_{2})\}, \nonumber\\
\rho'_{55}=&&\frac{1}{3}(\mu \overline{d}_{1} M_{1}+\overline{\mu}\overline{d}_{1}^{2}M_{2})\overline{q}_{2}^{2},\nonumber\\
\rho'_{99}=&&\frac{1}{3}(\mu \overline{d}_{2} M_{1}+\overline{\mu}\overline{d}_{2}^{2}M_{2})\overline{q}_{1}^{2},\\
\rho'_{15}=&&\rho_{51}^{'*}= \frac{1}{3}(\mu \sqrt{\overline{d}_{1}} M_{1}+\overline{\mu}\overline{d}_{1}M_{2})\overline{q}_{1}\overline{q}_{2}^{2},\nonumber\\
\rho'_{19}=&&\rho_{91}^{'*}= \frac{1}{3}(\mu \sqrt{\overline{d}_{2}} M_{1}+\overline{\mu}\overline{d}_{2}M_{2})\overline{q}_{1}^{2}\overline{q}_{2},\nonumber\\
\rho'_{59}=&&\rho_{95}^{'*}= \frac{1}{3}(\mu \sqrt{\overline{d}_{1} \overline{d}_{2}} M_{1}+\overline{\mu}\overline{d}_{1}\overline{d}_{2}M_{2})\overline{q}_{1}\overline{q}_{2},\nonumber
\end{eqnarray}
where $M_{1}= (1+\overline{d}_{1}^{2}+\overline{d}_{2}^{2})/3$ and $M_{2}=(1+\overline{d}_{1}+\overline{d}_{2})/3$.

Through the procedure of quantum teleportation, Bob obtains the state
\begin{eqnarray}
\label{eq22}
\rho_{\rm out}^{\rm EAM} &=& \frac{1}{M} \left[\begin{array}{ccc} \epsilon'_{00} & \epsilon'_{01} & \epsilon'_{02} \\ \epsilon'_{10} & \epsilon'_{11} & \epsilon'_{12}\\ \epsilon'_{20} &\epsilon'_{21} & \epsilon'_{22} \end{array}\right],
\end{eqnarray}
where $\epsilon'_{00}=(\rho'_{11}+\rho'_{55}+\rho'_{99})\alpha^{2}$,
$\epsilon'_{11}=(\rho'_{11}+\rho'_{55}+\rho'_{99})|\beta|^{2}$,
$\epsilon'_{22}=(\rho'_{11}+\rho'_{55}+\rho'_{99})|\delta|^{2}$,
$\epsilon'_{01}=\epsilon_{10}^{'*}= \alpha \beta^{*}(\rho'_{15}+\rho'_{59}+\rho'_{91})$,
$\epsilon'_{02}=\epsilon_{20}^{'*}= \alpha \delta^{*}(\rho'_{19}+\rho'_{51}+\rho'_{95})$,
$\epsilon'_{12}=\epsilon_{21}^{'*}= \beta \delta^{*}(\rho'_{15}+\rho'_{59}+\rho'_{91})$, and $M=\rho'_{11}+\rho'_{55}+\rho'_{99}$ is the normalization factor.
%\begin{eqnarray}
%\label{eq23}
%\epsilon'_{00}=&&(\rho'_{11}+\rho'_{55}+\rho'_{99})\alpha^{2}, \nonumber\\
%\epsilon'_{11}=&&(\rho'_{11}+\rho'_{55}+\rho'_{99})|\beta|^{2},\nonumber\\
%\epsilon'_{22}=&&(\rho'_{11}+\rho'_{55}+\rho'_{99})|\delta|^{2},\nonumber\\
%\epsilon'_{01}=&&\epsilon_{10}^{'*}= \alpha \beta^{*}(\rho'_{15}+\rho'_{59}+\rho'_{91}),\\
%\epsilon'_{02}=&&\epsilon_{20}^{'*}= \alpha \delta^{*}(\rho'_{19}+\rho'_{51}+\rho'_{95}),\nonumber\\
%\epsilon'_{12}=&&\epsilon_{21}^{'*}= \beta \delta^{*}(\rho'_{15}+\rho'_{59}+\rho'_{91}),\nonumber
%\end{eqnarray}

The average fidelity of EAM scheme could be written as 
\begin{equation}
\label{eq24}
\langle F\rangle^{\rm EAM}= \frac{1}{4}(2+B_{1}),
\end{equation}
where $B_{1}=(\rho'_{15}+\rho'_{19}+\rho'_{59}+\rho'_{51}+\rho'_{91}+\rho'_{95})/M$.
Again, we have to find the optimal strength of the post-QMRs. The details are given in Appendix \ref{app3} and the optimal strength of QMR is
\begin{equation}
\label{eq25}
q_{\rm opt}^{\rm EAM}=1-\Big(\overline{\mu}\overline{d}+\mu\sqrt{\overline{d}}\Big).
\end{equation}

Substituting Eq. (\ref{eq25}) into Eq. (\ref{eq24}), the optimal average fidelity of the EAM scheme is found to be
\begin{equation}
\label{eq26}
\langle F\rangle_{\rm opt}^{\rm EAM}=\frac{1}{2}+\frac{2[\overline{\mu}\sqrt{\overline{d}}(1+2\overline{d})+\mu(1+2\overline{d}^{2})](\overline{\mu}\sqrt{\overline{d}}+\mu)+\overline{\mu}\overline{d}+\mu+2\overline{d}^{2}}{2\Big\{[2(\mu\overline{d}+\overline{\mu})\overline{d}+1](\overline{\mu}\sqrt{\overline{d}}+\mu)^{2}+2(\overline{\mu}\overline{d}+\mu)+4 \overline{d}^{2}\Big\}}.
\end{equation}
The corresponding probability of success is
\begin{equation}
\label{27}
P_{\rm opt}^{\rm EAM}= \frac{1}{9}\overline{d}^{2}(\overline{\mu}\sqrt{\overline{d}}+\mu)^{2} \Big\{ [2(\mu\overline{d}+\overline{\mu})\overline{d}+1](\overline{\mu}\sqrt{\overline{d}}+\mu)^{2}+2(\overline{\mu}\overline{d}+\mu)+4 \overline{d}^{2} \Big\}.
\end{equation}

\begin{figure*}
 \includegraphics[width=0.8\textwidth]{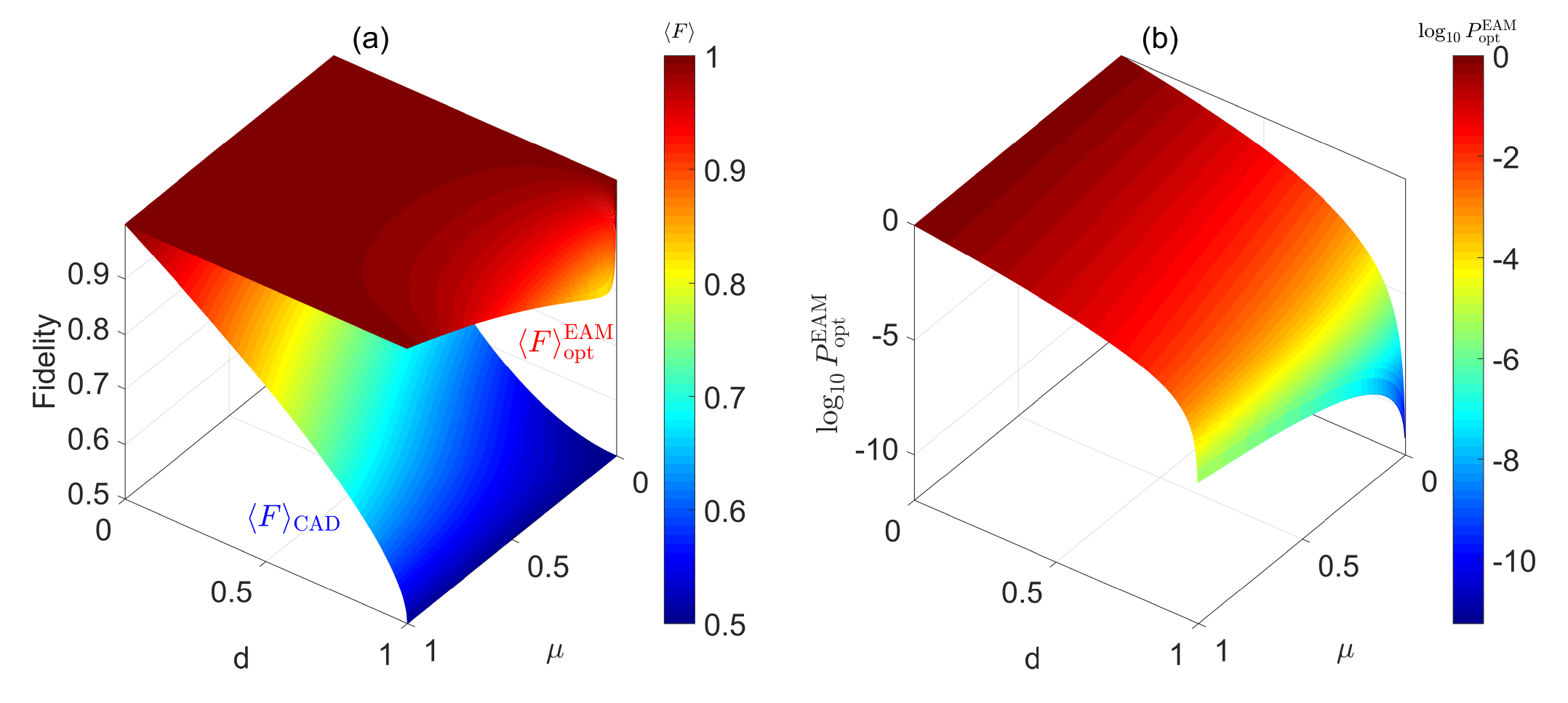}
\caption{(color online) (a) The average fidelities $\langle F\rangle_{\rm CAD}$,  $\langle F\rangle_{\rm opt}^{\rm EAM}$ and (b) the probability of success on a log scale $\log_{10}P_{\rm opt}^{\rm EAM}$ as a function of the correlated factor $\mu$ and noise strength $d$.}
\label{Fig4}       % Give a unique
\end{figure*}

In Fig. \ref{Fig4}, we analyze the behaviors of $\langle F\rangle_{\rm opt}^{\rm EAM}$ and $P_{\rm opt}^{\rm EAM}$ as a function of correlated factor $\mu$ and noise strength $d$. We also show the average fidelity $\langle F\rangle_{\rm CAD}$ in the case of CAD noise in Fig. \ref{Fig4}(a). We observe that the proposed EAM scheme significantly improves the fidelity compared to the original unprotected teleportation. However, we also find that the $\langle F\rangle_{\rm opt}^{\rm EAM}$ does not increase monotonically with the correlated factor $\mu$ for a fixed value of $d$ (e.g., $d\rightarrow1$). This can be understood as follows:
During the process of EAM, we have kept the measurement results of $k=0$, which correspond to the evolution of both $E_{00}$ and $A_{00}$, QMR cannot distinguish them exactly. Therefore, the fidelity can be restored to 1 by the operation of QMR only if AD noise ($\mu=0$) or FCAD noise ($\mu=1$) is involved. In other cases ($0<\mu<1$), the fidelity can be dramatically improved, but not fully recovered. Figure \ref{Fig4}(b) shows that the probability of success decreases with increasing $d$. The correlation effects also play a positive role in increasing the probability of success.

\section*{5 Discussions}
\label{sec5}
\begin{figure*}
 \includegraphics[width=0.8\textwidth]{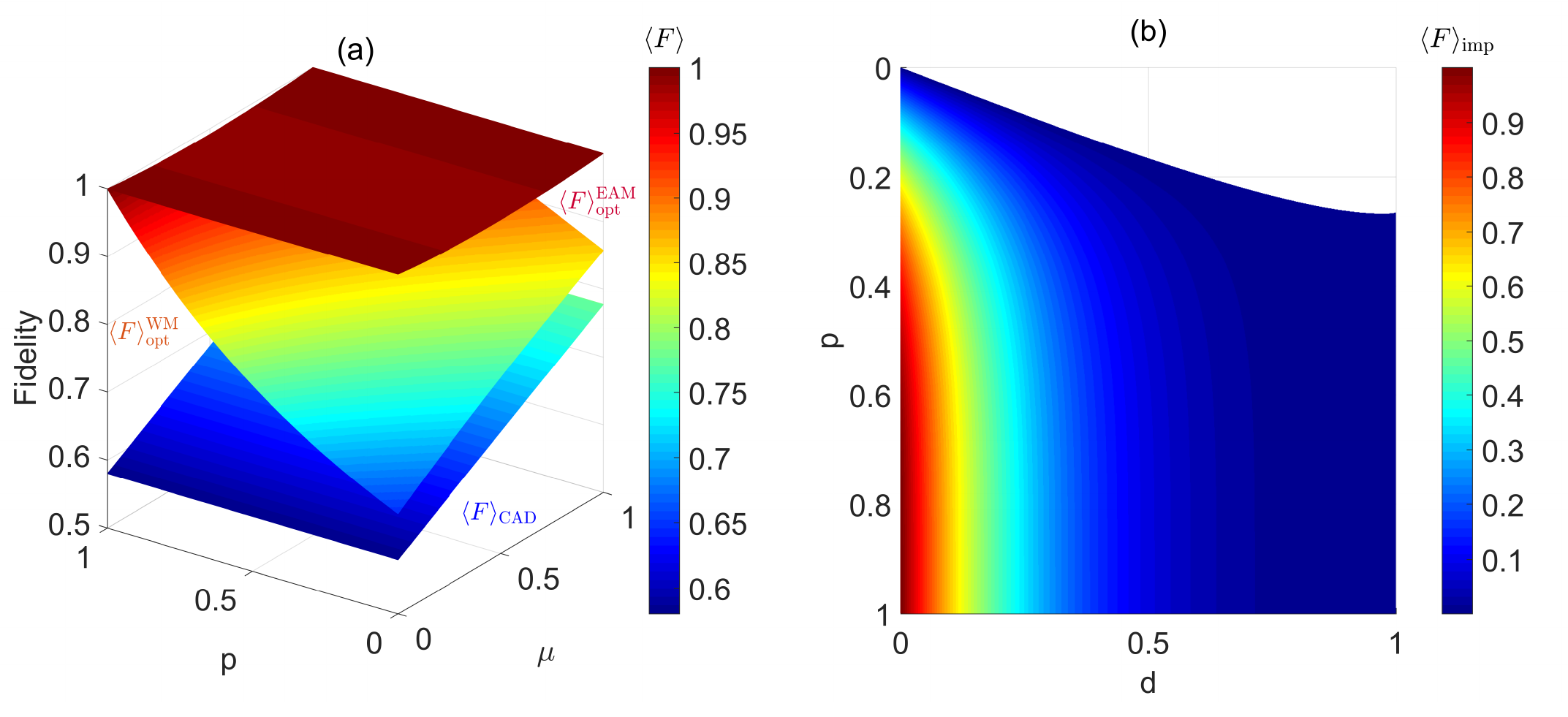}
 \caption{(color online) (a) The average fidelities $\langle F\rangle_{\rm CAD}$, $\langle F\rangle_{\rm opt}^{\rm WM}$, $\langle F\rangle_{\rm opt}^{\rm EAM}$ as a function of the correlated factor $\mu$ and WM strength $p$ with $d=0.6$. (b) The contour plot of the balanced fidelity improvement $ \langle F\rangle_{\rm imp}$ as a function of the WM strength $p$ and noise strength $d$ with $\mu=0.2$. The blank region indicates $ \langle F\rangle_{\rm imp}<0$.}
\label{Fig5}       % Give a unique
\end{figure*}

We have shown that both the WM and EAM schemes can significantly improve the fidelity.
It seems necessary to make some comparisons between them. Fig. \ref{Fig5}(a) shows the average fidelities $\langle F\rangle_{\rm CAD}$, $\langle F\rangle_{\rm opt}^{\rm WM}$, $\langle F\rangle_{\rm opt}^{\rm EAM}$ as a function of the correlated factor $\mu$ and the WM strength $p$ with $d=0.6$. We find that $\langle F\rangle_{\rm opt}^{\rm WM}$ is always larger than $\langle F\rangle_{\rm CAD}$, no matter what values $p$ and $\mu$ take. On the other hand, $\langle F\rangle_{\rm opt}^{\rm EAM}$ is always greater than or equal to $\langle F\rangle_{\rm opt}^{\rm WM}$ even if the strength of WM is close to 1. The above conclusion is also valid for other values of $d$. Note that these conclusions hold only when the private channels are noise-free. The case of private channels affected by AD noise will be discussed in Sec. 6.

Another aspect to note is that both the WM and EAM schemes are probabilistic.
As we discussed earlier, the greater improvement in fidelity comes at the cost of a lower probability of success in both schemes. Therefore, to make a fair comparison, we need to consider the effect of the probability of success at the same time. For this purpose, we introduce a quantity called balanced fidelity improvement
\begin{equation}
\langle F\rangle_{\rm imp}=\langle F\rangle_{\rm opt}^{\rm EAM}\times P_{\rm opt}^{\rm EAM}-\langle F\rangle_{\rm opt}^{\rm WM}\times P_{\rm opt}^{\rm WM}.
\end{equation}

The numerical simulation of the balanced fidelity improvement $ F_{\rm imp}$ as a function of the WM strength $p$ and noise strength $d$ is shown in Fig. \ref{Fig5}(b). It is interesting to note that although the EAM scheme consistently outperforms the WM strategy in terms of fidelity improvement, when considering the success probability comprehensively, $\langle F\rangle_{\rm imp}$ is not always greater than 0, particularly in regions where $p$ is relatively small. Therefore, the trade-off between the high fidelity and low success probability should be carefully balanced in realistic scenario. In some tasks, one may prefer to the fidelity as high as possible regardless of the low success probability (e.g., quantum teleportation and quantum state transfer), while in other tasks, the success probability is more important when the fidelity is enough (e.g., remote quantum state preparation).

%The numerical simulation of the balanced fidelity improvement $ F_{\rm imp}$ as a function of the correlated factor $\mu$ and noise strength $d$ is shown in Fig. \ref{Fig5}(b). To obtain greater fidelity, we selected $p=0.9$ as the strength of WM due to the limited fidelity improvement when $p$ is small. Notably, the balanced fidelity improvement $\langle F\rangle_{\rm imp}$ is always positive, which means that the EAM protocol is always better than the WM protocol even when we consider the probability of success. The reason can be solely attributed to the differences between WM and EAM.
%The EAM scheme is distinguished from WM by the fact that it is performed after the CAD noise, whereas WM is carried out before it. Essentially, WM only obtains information about the system, while EAM gathers information about both the system and the CAD noise. There can be no question that EAM collects more information than WM. Therefore, it can be deduced that the EAM approach will exceed the WM approach in enhancing the fidelity of teleportation.

\section*{6 Generalize to the noisy private channel}

\label{sec6}

\begin{figure*}
 \includegraphics[width=0.8\textwidth]{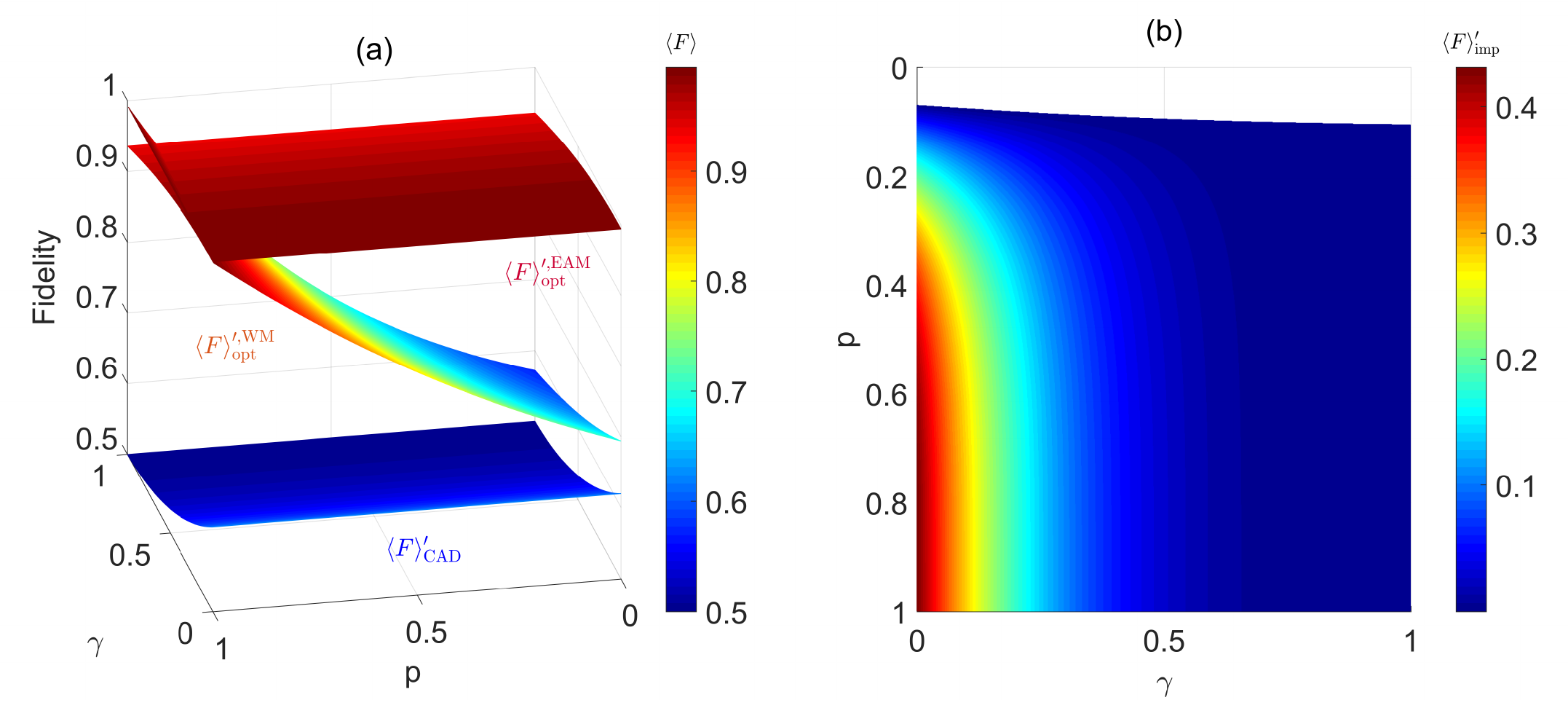}
 \caption{(color online) (a) The average fidelities $\langle F\rangle_{\rm CAD}^{\prime}$, $\langle F\rangle_{\rm opt}^{\prime,\rm WM}$, $\langle F\rangle_{\rm opt}^{\prime,\rm EAM}$ as a function of the correlated factor $\mu$ and WM strength $p$. (b) The contour plot of the balanced fidelity improvement $\langle F\rangle_{\rm imp}^{\prime}$ as a function of the WM strength $p$ and noise strength $\gamma$. The blank region indicates $\langle F\rangle_{\rm imp}^{\prime}<0$ The other parameters are $d=0.6$ and $\mu=0.2$.}
\label{Fig6}       % Give a unique
\end{figure*}

In above sections, we have confined our discussion to scenarios in which only the public channel is characterized as a CAD channel, while the private channels are assumed to be free of noise. Here, we turn to focus on the case where the private channels are also affected by AD noise. The characteristics of AD noise in the private channels are described by Eq. (\ref{eq2}), with the associated noise parameters denoted as $\gamma_1$ and $\gamma_2$. For the sake of simplicity, we again assume that both private channels are identical, such that $\gamma_1 = \gamma_2 = \gamma$. In this context, Alice and Bob will share the entangled state $\rho_{(23)}^{\prime,\rm WM}$ and $\rho_{(23)}^{\prime,\rm EAM}$ for WM and EAM schemes.
\begin{align}
\label{eq27}
\rho_{(23)}^{\prime,\rm WM}= &\mathcal{R}_{(23)}\Big\{\sum_{i,j=0}^{2}E_{ij}(\gamma)\mathcal{E}_{\rm CAD}\big[\mathcal{M}_{(23)}\big(|\Phi\rangle_{(23)}\langle\Phi|\big)\mathcal{M}_{(23)}^{\dagger}\big]E_{ij}^{\dagger}(\gamma)\Big\}\mathcal{R}_{(23)}^{\dagger},\\
\rho_{(23)}^{\prime,\rm EAM}=& \mathcal{R}_{(23)}\Big\{\sum_{i,j=0}^{2}E_{ij}(\gamma)\big[(1-\mu)E_{00}|\Phi\rangle_{(23)}\langle\Phi|E_{00}^{\dagger}\nonumber\\
&+\mu A_{00}|\Phi\rangle_{(23)}\langle\Phi|A_{00}^{\dagger}\big]E_{ij}^{\dagger}(\gamma)\Big\}\mathcal{R}_{(23)}^{\dagger}.
\label{eq28}
\end{align}
The fidelities $\langle F\rangle_{\rm opt}^{\prime,\rm WM}$ and $\langle F\rangle_{\rm opt}^{\prime,\rm EAM}$ can be obtained analytically, but the calculations and expressions are rather intricate and lack informativeness. For a better understanding of how the private channel noises affect the fidelity of quantum teleportation, we present the numerical results in Fig. \ref{Fig6}. 

Fig. \ref{Fig6}(a) shows the average fidelities $\langle F\rangle_{\rm CAD}^{\prime}$, $\langle F\rangle_{\rm opt}^{\prime,\rm WM}$ and $\langle F\rangle_{\rm opt}^{\prime,\rm EAM}$ as a function of the private noise parameter $\gamma$ and the WM strength $p$ with $d=0.6$ and $\mu=0.2$. We note that even when accounting for the noise in the private channel, both the WM and EAM schemes still demonstrate a significant improvement in fidelity. However, in contrast to Fig. \ref{Fig5}(a), the $\langle F\rangle_{\rm opt}^{\prime,\rm EAM}$ is not always larger than the $\langle F\rangle_{\rm opt}^{\prime,\rm WM}$, particularly when the strengths of private noise and WM are very strong. This indicates that the EAM cannot entirely eliminate the AD noise in the private channel, whereas the WM can achieve this suppression in the limit of $p\rightarrow1$. This phenomenon is readily comprehensible. As illustrated by Eq. (\ref{eq27}), the WM protocol effectively projects the entangled state into a `lethargic' state that exhibits substantial resilience to dissipative noise by implementing WM prior to the onset of noise, subsequently utilizing QMR solely in the final stage to restore the entangled state. Consequently, both CAD noise in the public channel and AD noise in the private channel can be effectively circumvented at the expense of extremely low success probability. 

The numerical simulation of the balanced fidelity improvement $\langle F\rangle_{\rm imp}^{\prime}$ as a function of the private noise parameter $\gamma$ and the WM strength $p$ is also shown in Fig. \ref{Fig6}(b). Although the EAM strategy is less effective than the WM scheme in the fidelity enhancement when $p$ and $\gamma$ approach 1, it exhibits greater robustness across most parameter ranges when considering both fidelity and success probability simultaneously.

\section*{7 Conclusions}

\label{sec7}
In summary, we propose two schemes to enhance the fidelity of qutrit teleportation in the CAD noise. The crucial ingredient for the WM scheme is the intentional inclusion of pre-WMs. This results in the initial entangled state collapsing partially towards the $|00\rangle$ state due to the state's inactivity in CAD noise. Due to the reversibility of WM, the fidelity of teleportation could be improved with the help of post-QMRs. The EAM scheme is based on the idea of feedback control, where an EAM is performed on the environment to detect the information and thereby restoring the fidelity by QMRs with a certain probability. Although they are probabilistic, the correlation effects also contribute positively to improving the probability of success in both schemes. 
Moreover, a comprehensive comparison indicates that the EAM scheme always surpasses the WM approach regarding fidelity in battling the CAD noise in the public channel. Even when taking into account the AD noise in the private channel, the EAM scheme outperforms the WM scheme in the vast majority of cases, unless the noise in the private channel is exceedingly severe.
Our findings indicate that the methods of WM and EAM possess the ability to handle not only AD noise but also effectively reduce CAD noise. Such a capability is crucial for various tasks of quantum information processing, especially when the transmission speed of information is too high to neglect the correlation effects.

%\section*{Appendices}
\begin{appendix}
\counterwithin{equation}{section}
\section{Kraus operators of CAD noise}
\label{app1}

The AD noise models the dissipative interaction between a quantum system and a zero-temperature bath. For the case of $V$-type qutrit system with levels $|0\rangle$, $|1\rangle$ and $|2\rangle$,
the time evolution is determined by the master equation
\begin{equation}
\dot{\rho}=-i[\hat{H},\rho]+\mathcal{L}\rho,
\end{equation}
where $\hat{H}$ is the Hamiltonian and $\mathcal{L}$ is the Lindblad superoperator. Under the Born-Markov approximation, the Lindblad
superoperator is
\begin{equation}
\mathcal{L}\rho=\frac{\gamma_{10}}{2}(2\sigma_{01}\rho\sigma_{10}-\sigma_{11}\rho-\rho\sigma_{11})+\frac{\gamma_{20}}{2}(2\sigma_{02}\rho\sigma_{20}-\sigma_{22}\rho-\rho\sigma_{22}),
\end{equation}
where $\gamma_{i0}$ are the spontaneous rates of the $i$th level. $\sigma_{ij}=|i\rangle\langle j|$ is a transition operator between $|i\rangle$ and $|j\rangle$. Note that we have considered the forbidden transition between two excited states $|1\rangle$ and $|2\rangle$. According to the Kraus decomposition $\rho(t)=\sum_{i}E_{i}\rho E_{i}^{\dagger}$, the Kraus operators of AD noise can be obtained as
\begin{eqnarray}
E_{0}=\left(\begin{array}{ccc}1 & 0 & 0 \\0 & \sqrt{\overline{d}_1} & 0 \\0 & 0 & \sqrt{\overline{d}_2}\end{array}\right),E_{1}=\left(\begin{array}{ccc}0 & \sqrt{d_1} & 0 \\0 & 0 & 0 \\0 & 0 & 0\end{array}\right),E_{2}=\left(\begin{array}{ccc}0 & 0 & \sqrt{d_2} \\0 & 0 & 0 \\0 & 0 & 0\end{array}\right),
\end{eqnarray}
with $d_1=1-\exp(-\gamma_{10}t)$, $d_2=1-\exp(-\gamma_{20}t)$. The Kraus operators of two-qutrit uncorrelated AD channel are tensor products of the single-qutrit Kraus operators $E_{ij}=E_{i}\otimes E_{j}$.

For characterizing the correlation effects of two qutrits successively passes through the same AD noise, it is necessary to calculate the correlated Lindblad equation of the two-qutrit state $\tilde{\rho}$. Note that $\tilde{\rho}\in C^{9\times9}$ is the total state of two qutrits. The correlated version of the Lindblad superoperator $\tilde{\mathcal{L}}$ can be expressed as \cite{Yeo2003}
\begin{equation}
\tilde{\mathcal{L}}\tilde{\rho}=\frac{\gamma_{10}}{2}\big(2\sigma_{01}^{\otimes2}\tilde{\rho}\sigma_{10}^{\otimes2}-\sigma_{11}^{\otimes2}\tilde{\rho}-\tilde{\rho}\sigma_{11}^{\otimes2}\big)+\frac{\gamma_{20}}{2}\big(2\sigma_{02}^{\otimes2}\tilde{\rho}\sigma_{20}^{\otimes2}-\sigma_{22}^{\otimes2}\tilde{\rho}-\tilde{\rho}\sigma_{22}^{\otimes2}\big),
\end{equation}
where $\sigma_{ij}^{\otimes2}=\sigma_{ij}\otimes\sigma_{ij}$. The above equation describes the physical process for which the decay emerges synchronously for both qutrits. The Kraus operators of the correlated part of the evolution of two-qutrit state are given as \cite{Jeong2019,Wang2022}
\begin{eqnarray}
A_{00}&=&\left(\begin{array}{cccc}{\rm I}_{4\times4} &  &  &  \\  & \sqrt{\overline{d}_1} &  &  \\  &   & {\rm I}_{3\times3} &   \\  &   &   & \sqrt{\overline{d}_2}\end{array}\right),
A_{11}=\left(\begin{array}{c|c|c} {\rm O}_{1\times4}  & \sqrt{d_1} & {\rm O}_{1\times4}   \\
\hline
 {\rm O}_{4\times4}   & {\rm O}_{4\times1}   &  {\rm O}_{4\times4} \\
\hline
  {\rm O}_{4\times4} & {\rm O}_{4\times1}   & {\rm O}_{4\times4}  \end{array}\right),\nonumber\\
A_{22}&=&\left(\begin{array}{c|c|c} {\rm O}_{1\times4}  &  {\rm O}_{1\times4}  & \sqrt{d_2}   \\
\hline
 {\rm O}_{4\times4}   & {\rm O}_{4\times4}   &  {\rm O}_{4\times1} \\
\hline
  {\rm O}_{4\times4} & {\rm O}_{4\times4}   & {\rm O}_{4\times1}  \end{array}\right)
\end{eqnarray}

\section{Derivation of Eq. (\ref{eqb14})}
\label{app2}
The most critical factor in preserving fidelity is to protect the initial shared entanglement between Alice and Bob from the CAD noise. An intuitive method for protecting entanglement is to ensure the finally entangled state as close as possible to the initially entangled state. From the perspective of pure state, one can find a optimal QMR by using the trick of ‘unraveling’ the excitation
into ‘jump’ and ‘no jump’. We assume the initially entangled state has the form as $(a|00\rangle+b|11\rangle+c|22\rangle)_{\rm s}$ and the environment is in a vacuum state $|\bf{0}\rangle_{\rm env}$. The procedures of WM, CAD and QMR can be expressed as:
\begin{eqnarray}
\label{eqb1}
&&\big(a|00\rangle+b|11\rangle+c|22\rangle\big)_{\rm s}|\bf{0}\rangle_{\rm env}\\%jump
\xrightarrow{\rm WM}&&\big(a|00\rangle+b\overline{p}_{1}|11\rangle+c\overline{p}_{2}|22\rangle\big)_{\rm s}|\bf{0}\rangle_{\rm env}\\
\label{eqb2}
\xrightarrow{\rm CAD}&&\overline{\mu}\big(a|00\rangle+b\overline{p}_{1} \overline{d}_{1}|11\rangle+c\overline{p}_{2} \overline{d}_{2}|22\rangle\big)_{\rm s}|\textbf{0}\rangle_{\rm env}\nonumber\\
\label{eqb3}
&&+\overline{\mu}b\overline{p}_{1}\sqrt{d_{1}\overline{d}_{1}}\big(|01\rangle+|10\rangle\big)_{\rm s}|\textbf{1}\rangle_{\rm env}+\overline{\mu}c\overline{p}_{2}\sqrt{d_{2}\overline{d}_{2}}\big(|02\rangle+|20\rangle\big)_{\rm s}|\textbf{1}\rangle_{\rm env}\nonumber\\
&&+\overline{\mu}\big(b\overline{p}_{1} d_{1}|00\rangle+c\overline{p}_{2} d_{2}|00\rangle\big)_{\rm s}|\textbf{2}\rangle_{\rm env}+\mu\big(b\overline{p}_{1} \sqrt{d_{1}}+c\overline{p}_{2} \sqrt{d_{2}}\big)|00\rangle_{\rm s}|\textbf{2}\rangle_{\rm env}\nonumber\\
&&+\mu\big(a|00\rangle+b\overline{p}_{1} \sqrt{\overline{d}_{1}}|11\rangle+c\overline{p}_{2} \sqrt{\overline{d}_{2}}|22\rangle\big)_{\rm s}|\textbf{0}\rangle_{\rm env}\\
\xrightarrow{\rm QMR}&&\Big[a\overline{q}_{1}\overline{q}_{2}|00\rangle+b\overline{p}_{1}\overline{q}_{2}\big(\overline{\mu}\overline{d}_{1}+\mu\sqrt{\overline{d}_{1}}\big)|11\rangle+c\overline{p}_{2}\overline{q}_{1}\big(\overline{\mu}\overline{d}_{2}+\mu\sqrt{\overline{d}_{2}}\big)|22\rangle\Big]_{\rm s}|\bf{0}\rangle_{\rm env}\nonumber\\
&&+\overline{\mu}b\overline{p}_{1}\sqrt{d_{1}\overline{d}_{1}}\overline{q}_{2}\sqrt{\overline{q}_1}\big(|01\rangle+|10\rangle\big)_{\rm s}|\textbf{1}\rangle_{\rm env}\nonumber\\
&&+\overline{\mu}c\overline{p}_{2}\sqrt{d_{2}\overline{d}_{2}}\overline{q}_{1}\sqrt{\overline{q}_2}\big(|02\rangle+|20\rangle\big)_{\rm s}|\bf{1}\rangle_{\rm env}\nonumber\\
&&+\Big[b\overline{p}_{1}\overline{q}_{1}\overline{q}_{2}\big(\overline{\mu}d_{1}+\mu\sqrt{d_{1}}\big)+c\overline{p}_{2}\overline{q}_{1}\overline{q}_{2}\big(\overline{\mu}d_{2}+\mu\sqrt{d_{2}}\big)\Big]|00\rangle_{\rm s}|\bf{2}\rangle_{\rm env},
\label{eqb4}
\end{eqnarray}
where $|\textbf{n}\rangle_{\rm env}$ denotes $n$ excitations in the environment. In order to ensure Eq. (B4) is close to Eq. (B1), one can choose 
\begin{eqnarray}
\label{eqb5}
\overline{q}_{\rm 1,opt}^{\rm WM}&=&\overline{p}_{1}\Big(\overline{\mu}\overline{d}_{1}+\mu \sqrt{\overline{d}_{1}}\Big),\\
\overline{q}_{\rm 2,opt}^{\rm WM}&=&\overline{p}_{2}\Big(\overline{\mu}\overline{d}_{2}+\mu \sqrt{\overline{d}_{2}}\Big).\nonumber
\end{eqnarray}
Then Eq. (B4) reduces to (up to a normalized factor $1/\sqrt{\overline{q}_{\rm 1,opt}^{\rm WM}\overline{q}_{\rm 2,opt}^{\rm WM}}$)
\begin{eqnarray}
\label{eqb6}
&&\Big(a|00\rangle+b|11\rangle+c|22\rangle\Big)_{\rm s}|\bf{0}\rangle_{\rm env}\\
&&+\overline{\mu}b\overline{p}_{1}\sqrt{\frac{d_{1}\overline{d}_{1}}{\overline{q}_{\rm 1,opt}^{\rm WM}}}\Big(|01\rangle+|10\rangle\Big)_{\rm s}|\textbf{1}\rangle_{\rm env}+\overline{\mu}c\overline{p}_{2}\sqrt{\frac{d_{2}\overline{d}_{2}}{\overline{q}_{\rm 2,opt}^{\rm WM}}}\Big(|02\rangle+|20\rangle\Big)_{\rm s}|\bf{1}\rangle_{\rm env}\nonumber\\
&&+\Big[b\overline{p}_{1}\big(\overline{\mu}d_{1}+\mu\sqrt{d_{1}}\big)+c\overline{p}_{2}\big(\overline{\mu}d_{2}+\mu\sqrt{d_{2}}\big)\Big]|00\rangle_{\rm s}|\bf{2}\rangle_{\rm env}.\nonumber
\end{eqnarray}
It is interesting to note that the state of (B6) becomes closer to (B1) with the increasing strength of WMs. Particularly, when $p_{1},p_{2}\rightarrow 1$, the last three terms in Eq. (B6) can be neglected and the initial state (B1) is exactly restored.

\section{Derivation of Eq. (\ref{eq25})}
\label{app3}
We start with the same initial state, but the sequence of the operations is CAD, EAM and QMR.
\begin{eqnarray}
\label{eqc1}
&&\big(a|00\rangle+b|11\rangle+c|22\rangle\big)_{\rm s}|\textbf{0}\rangle_{\rm env}\\%jump
\xrightarrow{\rm CAD}&&\overline{\mu}\big(a|00\rangle+b\overline{d}_{1}|11\rangle+c\overline{d}_{2}|22\rangle\big)_{\rm s}|\textbf{0}\rangle_{\rm env}\\
&&+\overline{\mu}b\sqrt{d_{1}\overline{d}_{1}}\big(|01\rangle+|10\rangle\big)_{\rm s}|\textbf{1}\rangle_{\rm env}+\overline{\mu}c\sqrt{d_{2}\overline{d}_{2}}\big(|02\rangle+|20\rangle\big)_{\rm s}|\textbf{1}\rangle_{\rm env}\nonumber\\
&&+\overline{\mu}\big(bd_{1}|00\rangle+cd_{2}|00\rangle\big)_{\rm s}|\textbf{2}\rangle_{\rm env}+\mu\big(b\sqrt{d_{1}}+c\sqrt{d_{2}}\big)|00\rangle_{\rm s}|\textbf{2}\rangle_{\rm env}\nonumber\\
&&+\mu\big(a|00\rangle+b\sqrt{\overline{d}_{1}}|11\rangle+c\sqrt{\overline{d}_{2}}|22\rangle\big)_{\rm s}|\textbf{0}\rangle_{\rm env}\nonumber\\
\xrightarrow{\rm EAM}&&\overline{\mu}\big(a|00\rangle+b\overline{d}_{1}|11\rangle+c\overline{d}_{2}|22\rangle\big)_{\rm s}|\textbf{0}\rangle_{\rm env}\\
&&+\mu\big(a|00\rangle+b\sqrt{\overline{d}_{1}}|11\rangle+c\sqrt{\overline{d}_{2}}|22\rangle\big)_{\rm s}|\textbf{0}\rangle_{\rm env}\nonumber\\
\xrightarrow{\rm QMR}&&\Big[a\overline{q}_{1}\overline{q}_{2}|00\rangle+b\overline{q}_{2}\big(\overline{\mu}\overline{d}_{1}+\mu\sqrt{\overline{d}_{1}}\big)|11\rangle+c\overline{q}_{1}\big(\overline{\mu}\overline{d}_{2}+\mu\sqrt{\overline{d}_{2}}\big)|22\rangle\Big]_{\rm s}|\bf{0}\rangle_{\rm env}.\nonumber\\
\label{eqb4}
\end{eqnarray}
It is easy to find that the choose of $\overline{q}_{1,\rm opt}^{\rm EAM}=\overline{\mu}\overline{d}_{1}+\mu \sqrt{\overline{d}_{1}}$ and $\overline{q}_{2,\rm opt}^{\rm EAM}= \overline{\mu}\overline{d}_{2}+\mu \sqrt{\overline{d}_{2}}$ can recover the initial state. If we assume $d_1=d_2=d$, then the optimal QMR of EAM scheme is obtained as
\begin{equation}
q_{\rm opt}^{\rm EAM}=1-\Big(\overline{\mu}\overline{d}+\mu\sqrt{\overline{d}}\Big).
\end{equation}

\end{appendix}

%%%%%%%%%%%%%%%%%%%%%%%%%%%%%%%%%%%%%%%%%%%%%%
%%                                          %%
%% Backmatter begins here                   %%
%%                                          %%
%%%%%%%%%%%%%%%%%%%%%%%%%%%%%%%%%%%%%%%%%%%%%%

\begin{backmatter}

\section*{Declarations}

\section*{Acknowledgements}%% if any
No applicable.

\section*{Funding}%% if any
X. Xiao is supported by National Natural Science Foundation of China under Grant Nos. 12265004, 11805040 and Jiangxi Provincial Natural Science Foundation under Grant No. 20242BAB26010. Y. L. Li is supported by National Natural Science Foundation of China under Grant Nos. 12365003 and Jiangxi Provincial Natural Science Foundation under Grant No. 20212ACB211004. T. X. Lu is supported by the National Natural Science Foundation of China under Grant No. 12205054.

%\section*{Abbreviations}%% if any
%AD, Amplitude Damping; CAD, Correlated Amplitude Damping; WM, Weak Measurement; EAM, Environment-assisted Measurement; QMR, Quantum Measurement Reversal.%

\section*{Availability of data and materials}%% if any
Not applicable.
\section*{Ethics approval and consent to participate}%% if any
Not applicable.
\section*{Competing interests}
The authors declare that they have no competing interests.

\section*{Consent for publication}%% if any
Not applicable.

\section*{Authors' contributions}
X.X. and Y.L.L. conceived and developed the idea, performed the calculations and analyzed the results. T.X.L. prepared the figures, discussed the results, and commented on the manuscript. Y.L.L. supervised the project. All authors reviewed and approved the final manuscript.

\end{backmatter}
\end{document}